\documentclass{aa}
\usepackage{graphicx}
\usepackage{txfonts}

\begin{document}

\title
{Massive elliptical galaxies in X-rays: the role of late gas accretion}

\author{Antonio Pipino\inst{1,2}, Daisuke Kawata\inst{2},
 Brad K. Gibson\inst{2} and Francesca Matteucci\inst{1}}
\institute{
$^1$Dipartimento di Astronomia, Universit\`a di Trieste,
    Via G.B. Tiepolo 11, I-34127, Trieste, Italy\\
$^2$Centre for Astrophysics and Supercomputing, Swinburne University of Technology, Hawthorn VIC 3122, Australia}
\offprints{A. Pipino, \email{antonio@ts.astro.it}}
\date{Accepted,
      Received }
      
\authorrunning{Pipino et al.}
\titlerunning{X-ray properties of massive elliptical galaxies}

\abstract{
We present a new chemical evolution model meant to be a first
step in the self-consistent study of both optical and X-ray properties of elliptical galaxies.
Detailed cooling and heating processes in the interstellar medium (ISM)
are taken into account using a mono-phase one-zone treatment which
allows a more reliable modelling of the galactic wind regime with respect to previous work.
The model successfully reproduces simultaneously the mass-metallicity, 
colour-magnitude, the $L_X - L_B$ and the $L_X - T$ relations,
as well as the observed trend of the [Mg/Fe] ratio as a function of $\sigma$, by adopting the prescriptions
of Pipino \& Matteucci (2004) for the gas infall and star formation timescales. 
We found that a late secondary accretion of gas from the environment plays a fundamental
role in driving the $L_X - L_B$ and $L_X - T$ relations and can explain their large observational scatter.
The \emph{iron discrepancy}, namely the too high predicted iron abundance in X-ray haloes of ellipticals
compared to observations, still persists. On the other hand, we predict [O/Fe] in the ISM which
is in good agreement with the most recent observations.
We suggest possible mechanisms acting on a galactic scale which may solve the iron discrepancy.
In particular, mixing of gas driven by AGNs may preserve the gas mass (and thus the X-ray luminosity)
while diluting the iron abundance. New predictions for the amounts of iron, oxygen and energy
ejected into the intracluster medium (ICM) are presented and we conclude
that Type Ia supernovae (SNe Ia) play a fundamental role in the ICM enrichment. 
SNe Ia activity, in fact, may power a
galactic wind lasting for a considerable amount of the galactic
lifetime, even in the case for which the efficiency of energy transfer
into the ISM per SN Ia event is less than unity.

\keywords{
galaxies: elliptical and lenticular, cD - galaxies: abundances - galaxies: evolution, X-rays: ISM, ICM.
}

}

\maketitle

\section{Introduction}

The most direct evidence for the presence of a non-negligible interstellar medium (ISM)
in elliptical galaxies comes from X-ray observations. The presence of
$\sim 10^{8} -10^{10} M_{\odot}$ of gas in a hot ($\sim 10^7$K) X-ray
emitting ($L_X \sim 10^{39} -10^{42} \rm erg\, s^{-1}$) phase,
was first discovered by the \emph{Einstein} satellite (e.g. Forman et al. 1979)
for ellipticals with a wide range of optical luminosities.
A correlation between X-ray ($L_X$) and optical ($L_B$) luminosities is observed with roughly $L_X \propto L_B^2$
(e.g. O'Sullivan et al. 2001a), although the scatter is quite large. Such
a trend can be explained by the fact that optically brighter
galaxies are also more gravitationally bound (Ciotti et al. 1991).
On the other hand, the X-ray emission from faint galaxies 
is dominated by discrete sources which have stellar origin
(Brown \& Bregman 2001)
and the above correlation becomes approximately $L_X \propto L_B$
(Canizares et al.\ 1987; O'Sullivan et al.\ 2001a).
In what follows, we focus on massive systems which follow the 
$L_X \propto L_B^2$ trend and whose X-ray data tell us something about
their ISM properties. Concerning the scatter of the $L_X-L_B$ relation, it is still not 
clear whether its origin can be traced to one of galactic environments
(see e.g. the recent review by Mathews \& Brighenti 2003 and references therein). 
However, ellipticals located in denser regions do seem to exhibit higher 
$L_X$ at a given $L_B$ with respect to their counterparts living
in low density environments (Helsdon et al.\ 2001; Matsushita 2001, but see Matsumoto et al.\ 1997
and O'Sullivan et al.\ 2001a).

As for clusters of galaxies, the X-ray properties of individual ellipticals 
follow an $L_X-T_X$ relation, although it is not a mere extrapolation
from the relation observed on the intergalactic scale (Matsushita et al.\ 2000),
because the so-called `entropy floor' seems to be more active on a galactic scale.
This fact can be easily understood in terms of the stellar feedback
still ongoing due to Type Ia supernovae (SNe Ia) and low mass stars as well as other gas-dynamical processes
(Matsumoto et al.\ 1997). A detailed study of these topics can be found in 
Kawata \& Gibson (2003b), who found that radiative cooling is an important 
driver of this relation, although it does lead to late star formation
(SF) and thus to galactic colours which are too blue with respect to
observations. The additional energy feedback provided by AGN may rectify 
this problem (e.g.\ Kawata \& Gibson 2004).

X-ray spectra also carry information pertaining to
the chemical composition of the interstellar gas. The
\emph{ASCA} satellite provided the first reliable measure of the iron
abundance in the hot ISM of ellipticals
(e.g. Awaki et al.\ 1994; Matsumoto et al.\ 1997). 
These observations led to the so-called
\emph{iron discrepancy} (Arimoto et al.\ 1997), the fact that 
the inferred iron abundance was much lower than the solar value, at odds not only with theoretical 
models for elliptical galaxies (Arimoto \& Yoshii 1987;
 Matteucci \& Tornamb\'e 1987), which predicted that their ISM should exhibit
[Fe/H]$>0$, but also with the mean metallicity of the stellar component inferred from optical spectra. 
Arimoto et al.\ (1997) analysed different possible sources of
this discrepancy such as dilution by intracluster medium (ICM), 
different binary populations (and thus SNe Ia rates)
with respect to spirals, the presence of dust, and the uncertainties involved
in the modelling of the Fe L lines. In fact,
Matsushita et al.\ (1997) showed that by taking into account systematic errors in ASCA data,
the iron abundance can possibly reach the solar value.  
This issue has been partly alleviated because central iron abundances
determined with \emph{ASCA} did not take into account temperature gradients 
(Buote \& Fabian 1998; Buote 1999), due to 
its limited spatial resolution.
When a multi-temperature plasma model is adopted,
the iron abundance in the hot ISM seems to increase from $\sim 0.2-0.4\, Fe_{\odot}$
at large radii, to solar (or slightly higher) inside a radius of $\sim 50$ kpc
(see Mathews \& Brighenti 2003).

Moreover, thanks to \emph{XMM} and \emph{Chandra} satellites, it is now possible
not only to obtain a more reliable measure of the iron abundance in the ISM, but also
to observe other chemical species like O, Mg, Si, N, S, C (e.g Buote et al.\ 2003; Sakelliou et al.\ 2002;
Gastaldello \& Molendi 2002; Xu et al.\ 2002; Matsushita et al.\ 2003).

Elliptical galaxies have well-studied optical
photo-chemical properties;
it is natural to consider whether the theoretical 
interpretations of these observable can simultaneously
provide a satisfactory explanation for their X-ray properties.
The X-ray spectra provide unique tools to study
these reservoirs of gas in which the evidence of the role played by 
recent SNe Ia activity and
mass loss, due to stellar winds, can be found.

Models for the chemical evolution of elliptical galaxies (e.g. Larson 1974; Matteucci $\&$ Tornamb\'e 1987; Arimoto
$\&$ Yoshii 1987) 
can successfully explain the mass-metallicity (e.g. Carollo et al.\
1993; Gonzalez 1993; Davies et al.\ 1993; Trager et al.\ 1998, 2000a, 200ob) and 
the colour-magnitude (CMR: e.g. Bower et al.\ 1992a) relations, as well as
the observed radial gradients in the line-strength indices and colours 
(e.g. Carollo et al.\ 1993; Peletier et al.\ 1990).
Under the assumption that the SF efficiency 
increases with the galactic mass 
(Tinsley \& Larson 1979; Matteucci 1994), and
that the accretion timescale is lower in higher mass systems,
Pipino \& Matteucci (2004, PM04 hereafter) were able to match simultaneously
the relations presented above and the magnesium overabundance with respect
to iron in the stellar component, which is observed to increase with galactic
mass (e.g. Faber et al.\ 1992; Carollo et
al.\ 1993; Davies et al.\ 1993; Worthey et al.\ 1992).

The aim of this paper is to improve the detailed
photo-chemical evolution code of PM04 with a new
treatment of the hot ISM, in order to present, for the first
time, a model able to predict in a \emph{self-consistent} manner
the properties of both stellar (i.e. optical) and hot gas (i.e. X-ray) components
of the flux emitted by elliptical galaxies.
We have modified the equations of chemical evolution
presented by PM04, by introducing a secondary gas inflow phase
and explicitly taking into account the mass flow during the galactic wind.
We next updated the energetics by implementing the radiative cooling of the ISM
following Kawata \& Gibson (2003a)
and adopting the procedure described by Kawata \& Gibson (2003b)
to generate an X-ray spectrum of the model galaxies.
In particular, we will show how the $L_X-L_B$ relation depends
critically on the interplay between the gas mass flow during the wind phase
and the small secondary accretion episode. 
The chemical code will be presented in Sec. 2, with particular emphasis on
the novelties related to this new energetic formulation, as well as the method
adopted in order to derive consistent X-ray properties. 
The models and the results will be shown and discussed in Secs. 3 - 6 and our conclusions will
be summarized in Sec. 7.

\section{The model}

\subsection{The chemical evolution code}

The adopted chemical evolution model is based on that
presented by PM04. In this particular case, however, we consider our model galaxies
as a single zone extending out to 10$R_{\rm eff}$, where $R_{\rm eff}$ is the effective radius (see
Table 1), with instantaneous mixing of gas. Moreover
we take explicitly into account a possible mass flow due to the galactic wind
and a possible secondary episode of gas accretion. Therefore, the
the equation of chemical evolution for a single element \emph{i} takes the following form:
\begin{eqnarray}
{d G_i (t) \over d t}  &= & -\psi (t) X_i (t)\,   \nonumber \\
& & +\int_{M_L}^{M_{B_m}} \psi (t-\tau_m) Q_{\rm mi}(t-\tau_m) \phi (m) dm\,  \nonumber \\
& & + A\int_{M_{B_m}}^{M_{B_M}} \phi (m) \left [ \int_{\rm\mu_{\rm min}}^{0.5} f(\mu) Q_{\rm mi}(t-\tau_{\rm m_2}) \psi (t-\tau_{\rm m_2}) d\mu \right ] dm \,  \nonumber \\
& &  +(1-A)\, \int_{\rm M_{\rm B_m}}^{M_{\rm B_M}} \psi (t-\tau_m) Q_{\rm mi}(t-\tau_m) \phi (m) dm\,  \nonumber \\
& &  +\int_{\rm M_{\rm B_M}}^{M_U} \psi (t-\tau_m) Q_{\rm mi}(t-\tau_m) \phi (m) dm\,  \nonumber \\
& & +({d G_i (t) \over d t})_{\rm infall}-W(t)X_i (t)
+({d G_i (t) \over d t})_{\rm acc}\,  ,	
\label{dgdt-eq}
\end{eqnarray}
where $G_i (t)$ is the normalized mass density
of the element \emph{i} at the time \emph{t} in the ISM. \rm
$X_i (t)$ 
is defined as the abundance by mass of the element \emph{i} (see
also Matteucci $\&$ Greggio, 1986,
Gibson, 1997, and PM04 for a comprehensive discussion of this equation). 
The initial mass function (IMF) has the usual form $\phi (m)\propto m^{-(1+1.35)}$, 
(Salpeter 1955), and it is normalized to unity in the mass interval $0.1 -100 M_{\odot}$.
By means of this equation we can calculate the evolution of 21 elemental species.
The total ISM normalized gas mass density at each timestep is then simply $G(t)=\Sigma_{i=1}^{21} G_i (t)$.
\rm
The term $ -\psi (t) X_i (t)$ gives the rate at which
the element \emph{i} is subtracted from the ISM owing to the SF process
(see Sec. 2.5 for details on the SF rate $\psi$), whereas
the integrals in the right-hand side of the equation give the 
rate at which the element \emph{i} is restored into the interstellar medium
as unprocessed or newly-synthesized element by low- and intermediate-mass 
stars, and/or SNe Ia and Type II supernovae (SNe II).
\rm In fact, one of the fundamental points upon which our model is based,
is the detailed calculation of the SN rates
as well as the use of finite stellar lifetimes
and detailed stellar nucleosynthesis (see Sec. 2.6).
For type Ia SNe rate, we assume a progenitor model 
made of a C-O white dwarf plus a red giant and the formalism of Greggio $\&$ Renzini (1983) and Matteucci $\&$ Greggio (1986):
\begin{equation}
R_{SNIa}=A\int^{M_{BM}}_{M_{Bm}}{\phi(M_B) \int^{0.5}_{\mu_m}{f(\mu)
\psi(t- \tau_{M_{2}})d \mu \, dM_{B}}}\, ,
\end{equation}
where $M_{\rm B}$ is the total mass of the
binary system, $M_{Bm}=3 M_{\odot}$ and $M_{BM}=16 M_{\odot}$ are the minimum
and maximum masses allowed for the adopted progenitor systems, respectively.
$\mu=M_2/M_{\rm B}$ is the mass fraction of the secondary, which
is assumed to follow the distribution law:
\begin{equation}
f(\mu)=2^{1+\gamma}(1+\gamma)\mu^\gamma\, .
\end{equation} 
Finally, $\mu_{m}$ is its minimum value and $\gamma=2$
The predicted type Ia SN explosion rate is constrained
to reproduce the present day observed value (Cappellaro et al., 1999), by
fixing the parameter $A$ in eq. (2). In particular, $A$ represents 
the fraction of binary systems in the IMF which are
able to give rise to SN Ia explosions. In the following we adopt $A=0.09$.

On the other hand, the type II SN rate is:
\begin{eqnarray}
R_{SNII} & = & (1-A)\int^{16}_{8}{\psi(t-\tau_m) \phi(m)dm}\nonumber \\
& + & \int^{M_U}_{16}{\psi(t-\tau_m) \phi(m)dm}\, ,
\end{eqnarray}
where the first integral accounts for the single stars in the 
range 8-16$M_{\odot}$, and $M_{U}$ 
is the upper mass limit in the IMF.
The stellar lifetimes $\tau_m$ are taken from Padovani \& Matteucci (1993).
The knowledge of the stellar lifetimes and the star formation history
allow us to calculate the evolution of the stellar mass  
as given by:
$${d M_* (t) \over d t} = \psi (t) - \int_{m(t)}^{M_{U}} \psi (t-\tau_m) (m-m_{rem}) \phi (m) dm \, ,$$
where $m(t)$ is the smallest mass dying at the time $t$ and $m_{rem}$ is the mass
of the \emph{remnant} of a star of initial mass \emph{m}.
Therefore we can easily evaluate the number of stars still alive,
their distribution in mass as well as their distribution as a function of the metallicity.
In fact, for each simple stellar population (SSP) created, we store information
on the chemical composition of the gas out of which this SSP has formed.
\rm

The initial galactic infall phase enters the equation via the term:
\begin{equation}
({d G_i (t) \over d t})_{\rm infall}= X_{\rm i,infall} C e^{-{t \over \tau}}\, ,
\end{equation}
where $X_{\rm i,infall}$ describes the chemical composition of the accreted gas, assumed to be primordial.
$C$ is the normalisation constant of the infall law, obtained by integrating this law over time and
requiring that $M_{\rm lum}$ has been accreted at $t_{\rm gw}$ (see PM04). 
After PM04,
the initial collapse becomes faster as the galactic mass increases.

At variance with previous semi-analytical works 
(PM04; Pipino et al.\ 2002, P02 hereafter, but see also
Arimoto 1989; Gibson \& Matteucci 1997), in which all
the gas present at the time of the galactic wind is ejected into the IGM/ICM,
in this model we take into account the possibility that only a fraction
of the gas is able to escape the potential well. 
In order to do that we introduce the mass flow rate $W(t)$ in the 
chemical evolution equation. The functional form 
of $W(t)$ as well as its dependence on the energy content are given in Sec. 2.3.
 
Finally, in order to explore the possible interaction between
the pristine gas of the IGM/ICM and the hot ISM of the galactic models, 
we introduced an additional accretion episode, by means of the term
$({d G_i (t)/ d t})_{\rm acc}$ (see Sec. 2.4).

The detailed photometric evolution for our model elliptical galaxies
is obtained by applying the spectro-photometric code by Jimenez et al.\ (1998, see also PM04)
to our SSPs.

\subsection{Energy evolution}

 The chemical code presented in this paper features a new
self-consistent energetic treatment which supersedes the previous
one adopted by PM04. In particular,
calling $u_{\rm th}$ the thermal energy per unit mass,
the equation governing the energy evolution of the gas is:
\begin{equation}
{d u_{\rm th}\over dt}={\Gamma_{\rm SN}\over M_{\rm gas}}-{n_H^2\Lambda(Z,T)\over \rho_{\rm gas}}+
{1 \over  M_{\rm gas}}{d M_w \over d t}\cdot(u_{\rm vir} - u_{\rm th})\, ,
\label{duth-eq}
\end{equation}
where the first term on the right-hand side is the heating rate per unit mass (Sec. 2.2.2),
while the second is the cooling rate per unit mass (Sec. 2.2.1).
\rm
The last term represents the variation in the energy content due to the stellar mass loss
(injected into the ISM and being at the virial temperature of the system, see eq. \ref{ktvir-eq}).
This is a clearly simplified case with respect to those
presented in other works (e.g. Ciotti et al. 1991, Loewenstein \& Matthews, 1987,
Hattori et al. 1987). Nevertheless, because we cannot treat gas dynamics, we
believe that eq. (\ref{duth-eq}) is a good starting point for a detailed inclusion
of stellar feedback in a chemical evolution model.
\rm
The temperature of the ISM is related to the energy per unit mass
by means of:
\begin{equation}
u_{\rm th}={3\over 2\mu m_p}kT
\end{equation}
Equation (\ref{duth-eq}) has the form:
\begin{equation}
{d u_{\rm th}\over dt}=f(u_{\rm th})\, .
\end{equation}
Therefore, in order to solve equation (\ref{duth-eq}) 
for each time-step, we employ an iterative
process. In particular, we implemented a routine based on Brent's Method (see Press et al.\ 1986)
in the chemical code.

This treatment is needed in order to study the evolution of the thermal budget
of the ISM itself. The previous formalism, based on the thermal energy 
of the supernova remnant (SNR), is now part of the new procedure, so that we
can distinguish between the energy content of the gas and the energy
trapped in the hot bubble of the SNR. The latter is eventually restored 
to the ISM when the SNR merges with the surrounding medium. 
Moreover, the inclusion
of a consistent calculation of the cooling functions allows us
to compute the X-ray flux emitted by the model galaxies (see Sec. 2.7).

Equation (\ref{duth-eq}) relies on the assumption that each gas particle
shares the same energy, whether it is gas being used to form stars
or gas escaping in the galactic wind.
This limitation is due to the single-phase nature adopted
for the ISM. A more sophisticated treatment would require a two-phase model
in which mass exchange is allowed by evaporation
(cold phase$\rightarrow$hot phase) and
cooling (hot$\rightarrow$cold) (Harfst, Theis \& Hensler 2003).
The model presented in this paper, however,
is the first step in modeling both the X-ray and optical properties in 
a self-consistent manner.

The relation:
\begin{equation}
kT_{\rm vir}= {\mu m_p \over 2}{G M_{\rm vir}\over R_{\rm vir} }\, ,
\label{ktvir-eq}
\end{equation}
(e.g. Romano et al.\ 2002) gives us the initial value 
for the energy per unit mass; $R_{\rm vir}$ is the virial
radius for a spherical system of mass $M_{\rm vir}$, namely the mass of luminous
and dark matter inside 10 $R_{\rm eff}$.

In the case of secondary accretion, we take into account that
the infalling gas has a mean energy assumed to be equal to the virial temperature
(according to eq. \ref{ktvir-eq}), 
therefore influencing the final ISM temperature. 
This can be easily done by adding the term ${1 \over  M_{\rm gas}}{d M_{acc} \over d t}\cdot (u_{\rm vir}-u_{\rm th})$
in the right-hand side of eq. (6).

We limit our calculation to temperatures above $T=10^4$ K, 
since the cooling functions become more complex
at lower temperatures (e.g.\ Omukai 2000).
This condition, however, did not affect the final results, since in all 
the runs completed the gas 
temperature was always higher than this minimum value.

\subsubsection{The cooling term}

After Kawata \& Gibson (2003a, where we address the reader for a more detailed description),
we evaluate the cooling rate $\Lambda(Z,T)$
as a function of gas density and metallicity by means of
the cooling curves computed by MAPPINGS III (Sutherland \& Dopita 1993).
In particular, we used a bi-linear interpolation
in [Fe/H] and T in the range [$-3$,0.5] in metallicity and [$10^4$K, $10^9$K]
in temperature. Since our models span a wider range in [Fe/H] during their 
evolution, we assumed that the cooling rate of the gas with metallicity
[Fe/H]$<-3$ was the same as for 
gas with [Fe/H]$=-3$ in order to avoid uncertainties due to
extrapolation.
A similar assumption has been made for gas with metallicity [Fe/H]$>0.5$.

\subsubsection{The heating term}

The term describing the heating of the gas in equation (\ref{duth-eq}) 
is defined as:

\begin{equation}
\Gamma_{\rm SN}= \epsilon_{\rm SN Ia}E_0 R_{\rm SN Ia} 
+ \epsilon_{\rm SN II}E_0 R_{\rm SN II}\, ,
\end{equation}
where $E_0=10^{51}$erg is the assumed initial energy budget for the SNR,
$\epsilon_{\rm SN Ia}$ ($\epsilon_{\rm SN II}$) and $R_{\rm SN Ia}$ 
($R_{\rm SN II}$)
are the efficiency of energy transfer and the explosion rate for SNe Ia 
(SNe II), respectively (see eqs. 2 and 4).

At variance with previous works (P02; PM04),
here we adopt a parametric value for both 
$\epsilon_{\rm SN Ia}$ and $\epsilon_{\rm SN II}$.
This allows a more flexible treatment of
the heating term in order to test different scenarios. 
In particular, before the galactic wind, we focus on three different cases, namely
$\epsilon_{\rm SN Ia}=\epsilon_{\rm SN II}=$0.01,
$\epsilon_{\rm SN Ia}=\epsilon_{\rm SN II}=$0.1 and
$\epsilon_{\rm SN Ia}=\epsilon_{\rm SN II}=$1. 
We note in passing that our fiducial case 
($\epsilon_{\rm SN Ia}=\epsilon_{\rm SN II}=$0.1)  
is in good agreement with the results of Thornton et al.\ (1998), who
modelled SNR evolution within a range of different environments
finding that the fiducial value adopted here can be considered as typical.
In the following, the fiducial case has been adopted
unless otherwise noted. 
Furthermore, we have done preliminary calculations
for typical values of metallicity, density and temperature relevant to
our models, based on the Cioffi et al.\ (1988) formalism
(see also Gibson 1997). In particular, 
considering that a SNR merges with the ISM
(and hence eventually shares its energy with the surrounding medium) 
when its expansion velocity equals the gas sound speed, 
we found that the typical fraction
of the initial energy budget available for the ISM is only a few percent, again in agreement
with our fiducial case.

\rm
After the galactic wind has developed, the gas content of the galaxy
decreases quite sharply, whereas the temperature is almost constant
or increases. When we take into account these new conditions in the Cioffi et al.\ (1988)
equations, we have an increase in $\epsilon_{\rm SN Ia}$
up to values as high as 0.5-0.7 (see Table 1). This is the most natural way to 
inject energy and drive a wind in our model, but, in a more
general treatment (part of) this energy could be also 
provided by an AGN. 
\rm

A more detailed and self-consistent treatment of the SNR evolution
as a function of gas metallicity and temperature
will be included in a forthcoming paper.
\rm
A more general form for the heating term would request
the inclusion of the energy input from
stellar winds (see e.g. Loewenstein \& Mathews, 1987).
Its typical value for massive stars 
is $\sim 1\%$ of the assumed SNII energy budget (Mathews \& Baker, 1971).
Therefore the inclusion of stellar winds as a heating source has a negligible impact on
the final results (Bradamante et
al. 1998; Gibson 1996).
\rm

\subsubsection{The development of the galactic wind}

We define the gas binding energy $E_{\rm Bin}$ (see PM04
for its analytic definition) as the work required 
to carry a gas particle out to 10$R_{\rm eff}$. According to PM04,
we assume that the dark matter (DM) is distributed in a diffuse halo
ten times more massive than the baryonic component of the galaxy
with a scale radius $R_{\rm dark}= 10 R_{\rm eff}$. 
The DM profile is taken from Bertin et al.\ (1992).

In this paper we define the time when the galactic wind onsets
($t_{\rm gw}$) as the solution of
the following equation:
\begin{equation}
E_{\rm th}(t_{\rm gw}) = E_{\rm Bin}(t_{\rm gw})\, ,
\end{equation}
here the thermal energy of the gas is simply described by
\begin{equation}
E_{\rm th}=u_{\rm th}M_{\rm gas}\, .
\end{equation}
For every time-step in which the thermal energy is greater
than the binding energy, we set the energy loss
in the wind as the value given by $\Delta E = E_{\rm th}-E_{\rm Bin}$, whereas
in previous works (P02; PM04) all the thermal
energy was allowed to escape.

Before introducing the modelling of the SF history and 
the possible exchanges of matter with the surrounding
medium at $t>t_{\rm gw}$, we spend a few words 
on the differences between this new treatment for the energy evolution and our previous works. 
Since the heating term is not dramatically different from the energy source
powering the winds in P02 and PM04 models, which
featured an average SN efficiency of $\sim 20\%$, the most relevant difference
is the cooling of the ISM. This results in a lower effective contribution
by a single SN to the energetic budget and, thus, 
delayed galactic winds with respect
to previous models. Moreover, owing to the one-zone nature of this model,
the adopted galactic radius $R$ becomes an important parameter in determining
the cooling term, because the latter scales as $\rho_{\rm gas}^2$ (and thus as $R^{-6}$).
The influence of the adopted radius on the ISM X-ray properties will be discussed in Sec. 5.2.
The detailed energetic treatment will provide us with the tools to
evaluate the amount of gas which has enough energy to escape from the galactic potential well
and to control whether the conditions for the SF hold during 
the different stages of the galactic evolution.

\subsection{The mass flow rate}

The mass flow rate in the right-hand side
of equation (\ref{dgdt-eq}) is defined as:
\begin{equation}
W(t)={\Delta M \over\Delta t }{1\over M_{\rm lum}}\, .
\end{equation}
Given the assumptions leading to eq. (6) and the fact that
the amount of energy which can drive the wind is
$\Delta E$, the mass of gas which can blow away
from the galactic potential well at each time-step is given
by:
\begin{equation}
{\Delta M \over M_{\rm gas}}={\Delta E \over E_{\rm th}}\, .
\end{equation}
Therefore, the energy feedback enters in the equation
of the chemical evolution by means of eq. (13).

\subsection{Second Infall Episode}

Theoretical models based on the Cold Dark Matter paradigm predict that massive halos
keep accreting either dark or baryonic matter all their life
(e.g. Voit et al.\ 2003). Since our aim is to reproduce the properties of very bright ellipticals, we further
modified our chemical evolution code by adding the possibility of a late time accretion
of gas. In this way we partly relaxed the somewhat rough approximation that the infall suddenly stops 
at the occurrence of the galactic wind. It is likely, in fact, that even in the wind phase,
small amounts of gas can flow inward, e.g. in the presence of an inhomogeneous ISM.
On the other hand, the optical properties of elliptical galaxies require
that the main accretion episode must have been very fast compared to the
galactic lifetime and that most of the galaxy has to be assembled during 
this stage (e.g. PM04; Ferreras \& Silk 2003).
Therefore, in the following we shall refer only to the main accreting episode as the \emph{infall}.
In order to handle easily the parameters governing the secondary inflow,  
we chose to model independently these two accretion episodes and we adopted
the following expression for the second one:
\begin{equation}
({d G_i (t) \over d t})_{\rm acc}= X_{\rm i,acc} {M_{\rm acc} \over M_{\rm \rm lum} t_0}\, ,
\end{equation}
namely we assume that the model galaxy accretes a gas mass $M_{\rm acc}$ in a
uniform way all along its lifetime ($t_0$).
This choice can be motivated by the fact that the secondary accretion
is completely different from the fast initial infall. Eq. (15), in fact,
models in a simple way interactions with the environment or random accretions of a singular gas cloud
occurring on long timescales, within a one-zone chemical model with 
instantaneous mixing.
Also, in this case the accreted gas is assumed to be primordial and its
temperature is given by equation (\ref{ktvir-eq}).
In Sec. 5 we show the importance of this secondary accretion episode,
in governing the late time behaviour of the galaxies.

\subsection{Star formation history}

The SF rate (SFR) is assumed to be proportional to the gas density,
according to:
\begin{equation}
\psi (t)= \nu\cdot \rho_{\rm gas} (t)/\rho_{\rm gas,0} \, .
\end{equation} 
The SF efficiency 
$\nu$ is taken to be an increasing function of the galactic mass following the prescriptions
of the PM04 best model.

An improvement of this model with respect to PM04 is that
we can keep the gas temperature and its cooling
time under control. These quantities allow us to check in a self-consistent
manner whether the SF process might continue or not.
In particular, 
the SF takes place if the condition
$t_{\rm cool}\le  t_{\rm ff}$ is satisfied by the ISM gas. The cooling timescale
is defined as:
\begin{equation}
t_{\rm cool}= {E_{\rm th}\over {dE_{\rm th}\over dt}}={3\rho_{\rm gas}\over 2\mu m_p}{kT\over n_H^2 \Lambda(Z,T)}\, ,
\end{equation}
While the free-fall time is:
\begin{equation}
t_{\rm ff}= ({3\pi \over 32 G \rho_{\rm gas}})^{0.5}\, .
\end{equation}

As it can be seen from Fig. 1, the condition is satisfied at the
beginning of the SF process. Since then, 
because elliptical galaxies cannot stop or regulate the SF process (Elmegreen
1999), the model galaxies undergo a rapid and very efficient conversion of gas into stars
until the galactic wind occurs. On the other hand,
after the galactic wind the gas is so hot and rarefied
that the $t_{\rm cool}\le  t_{\rm ff}$ condition cannot hold anymore, therefore leading to 
a sudden stop in the SF, in agreement with previous works (PM04). 

Finally, when the galactic wind stops, late SF episodes are still inhibited by the high
temperature and very low density of the ISM, which imply $t_{\rm cool}> t_{\rm ff}$.
It is worth noting,
however, that the chemical evolution equation presented here holds
even in the case in which wind and SF are present at the same time.

\begin{figure}
\includegraphics[width=\hsize]{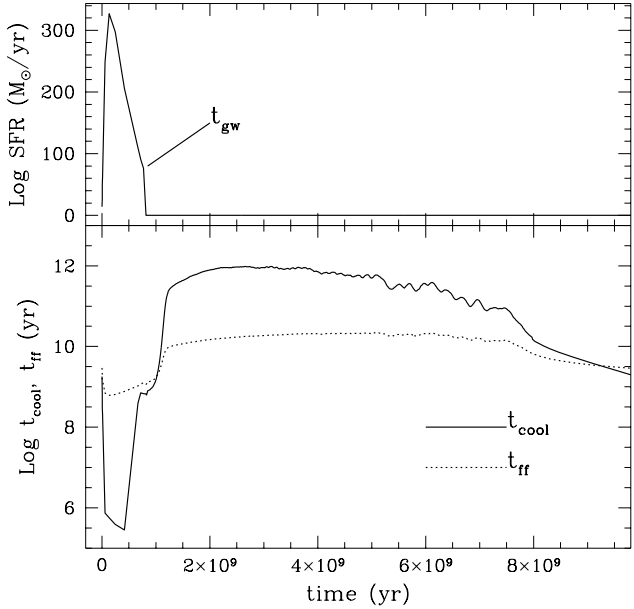}
\caption{Star formation rate (upper panel), $t_{\rm cool},\,t_{\rm ff}$
(lower panel) versus time for model La1 (see text and Table 1 for details).}
\end{figure}

\subsection{Stellar Yields}

In order to follow the evolution of the chemical elements,
we utilize the following nucleosynthesis prescriptions.
\begin{enumerate}
\item
Yields of van den Hoek $\&$ Groenewegen (1997, as a function of metallicity) for single low 
and intermediate mass stars ($0.8 \le M/M_{\odot} \le 8$).
\item
Yields by Nomoto et al.\ (1997) for SNe Ia.
\item
Yields of Thielemann et al.\ (1996),
which refer to the solar chemical composition, for massive stars ($M >8 M_{\odot}$). 
\end{enumerate}

\subsection{X-ray properties}

In order to study the X-ray emission from model galaxies
in a detailed manner, taking into account the line emissions
and the energy range of the actual observed spectra,
we applied the procedure described by
Kawata \& Gibson (2003b) to our chemical code, which we briefly
outline here.
In particular, using the elemental abundances
in the ISM and the gas temperature and density as provided by the chemical code, 
we create a spectrum in the energy range 0.1-20 keV, with 1800 bins, by means of the \bf vmekal \rm
plasma model (Mewe et al.\ 1985; Mewe et al.\ 1986; Kaastra 1992;
Liedahl et al.\ 1995). Then we generate a fake observed spectrum by convolving
the spectrum with the epn\_ff20\_sY9\_thin.rmf response function
for the EPN detector on-board the \emph{XMM-Newton} satellite in XSPEC ver.11.1.0 environment. The exposure time
is 40 ks and neither absorption component nor background were taken into account.
The spectrum is rebinned in order to have at least 25 counts per bin, and,
finally, fitted with XSPEC \bf vmekal \rm model in the same energy range of the different observations. We stress
that the abundance obtained as free parameters in the fitting procedure,
are in good agreement with the output of the chemical code.

A detailed analysis of this procedure, as well as
a comparison with other works in which the X-ray emission is modelled
as simply proportional to $\rho^2 T^{1/2}$, can be found in Kawata \& Gibson (2003b).
At variance with that paper, however, because we are presenting a one-zone model,
we will not take into account aperture effects.

\section{Galactic models}

We run models for elliptical galaxies with baryonic mass 
$M_{\rm lum}=10^{11}$ (models labelled as L) and $M_{\rm lum}=10^{12}M_{\odot}$ (models labelled as H).
$M_{\rm lum}$ is the `nominal' mass of the object, i.e. the mass of gas accreted and possibly turned into stars
during its active evolution (recall that we stop the infall when 
$t\sim t_{\rm gw}$). 
Furthermore, we name the models `0', `a1' or `a10' 
if they have no secondary accretion,
accrete 1 per cent or 10 per cent of $M_{\rm lum}$, i.e. $M_{\rm acc}$=0, 
0.01$M_{\rm lum}$ and 0.1$M_{\rm lum}$, 
during their whole evolution, respectively.
Models called HSN1 and HSN100 are similar to model Ha1, except for the SN efficiency (1 and 100 per cent, respectively).  
Since the secondary accretion episode implies a very mild accretion
rate, the SF history, and hence the
photochemical properties are not affected by it; only the late behaviour 
of the ISM properties depends strongly on the secondary accretion, therefore
affecting the X-ray luminosity and temperature.
A summary of the input parameters are shown in Table 1, where
model name (Col. 1), mass (Col. 2),
effective radius (Col. 3), SF efficiency (Col. 4),
infall timescale (Col. 5), SNe efficiency (Col. 6-8), accreted mass in the secondary
episode (Col. 10) as well as the times at which the galactic wind starts and stops
(Col. 9 and 11, respectively) are given.

\begin{table*}
\centering
\begin{minipage}{120mm}
\scriptsize
\begin{flushleft}
\caption[]{Model parameters}
\begin{tabular}{llllllllllll}
\hline
\hline
Model&$M_{\rm lum}$ 	&$R_{\rm eff}$ &  $\nu$ 	    & $\tau$
 &$\epsilon_{\rm SNII}$ & $\epsilon_{\rm SNIa}$ &$\epsilon_{\rm SNIa}$ &$t_{\rm gw}$& $M_{\rm acc}/M_{\rm lum}$&wind stop \\
&({$M_{\odot}$}) & ({kpc})  &  ({$\rm Gyr^{-1}$})& (Gyr)& &($t\le t_{\rm gw}$) &($t> t_{\rm gw}$) &(Gyr)   & &(Gyr) \\
\hline
L0&$10^{11}$       & 3        & 15             & 0.3 &0.10  &  0.10 & 0.50    &0.82 & 0&never \\
La1&$10^{11}$       & 3        & 15             & 0.3  &0.10 &   0.10 & 0.50    & 0.82 & 0.01&7.8 \\
La10&$10^{11}$       & 3        & 15             & 0.3 &0.10  &  0.10 & 0.50     &  0.82 &0.1&3.9  \\
H0&$10^{12}$       & 10       & 25             & 0.2  &0.10 &   0.10 &0.70     & 0.49 &0 &5.2	\\
Ha1&$10^{12}$       & 10       & 25             & 0.2  &0.10 &  0.10 &0.70     &  0.49 &0.01&3.7	\\
Ha10&$10^{12}$       & 10       & 25             & 0.2  &0.10 &  0.10 & 0.70    &  0.49 & 0.1&2.6	\\
HSN1&$10^{12}$       & 10       & 25             & 0.2  &0.01 &   0.01 & -    & no wind & 0.1& -	\\
HSN100&$10^{12}$       & 10       & 25             & 0.2  &1.00 &   1.00 & 1.00    & 0.13 & 0.1& never	\\
\hline
\end{tabular}
\end{flushleft}
\end{minipage}
\end{table*}

\section{The stellar component - photochemical properties of the models}

As it can be seen from Table 1, in this paper we follow the prescriptions
given by PM04 in order to reproduce the majority of photochemical observables for elliptical
galaxies, namely that the SF efficiency has to increase with the galactic mass, 
whereas the infall timescale must decrease with luminosity.
It is worth noting
that the values for $\nu$ and $\tau$ are very close to those presented by PM04.
Therefore we expect that our model 
succeeds in reproducing the mass-metallicity 
and the [Mg/Fe]-$\sigma$ relations
as well as the CMRs. We recall here that the observations of the increase of line-strength indices (such as 
Mg$_2$ and $\langle$Fe$\rangle$)
 with galactic velocity dispersion (e.g. Carollo et al.\
1993; Gonzalez 1993; Davies et al.\ 1993; Trager et al.\ 1998, 2000) 
is generally interpreted as
a metallicity sequence in which the stars of more massive galaxies are richer in metals
with respect to the stellar component of less massive objects. 
This metallicity sequence also explains the slope and tightness of the 
observed CMR (e.g. Kodama \& Arimoto 1997). On the other hand, the
observed increase in the mean stellar [Mg/Fe]
in the central region for ellipticals with galactic velocity dispersion
(e.g. Faber et al.\ 1992; Carollo et al.\ 1993; Davies et al.\ 1993; 
Worthey et al.\ 1992, but see Proctor et al.\ 2004),
tells us that the most massive galaxies must have formed in a shorter 
timescale compared to the less massive ones (time-delay model, 
see Matteucci 2001). 
We address the reader interested in a detailed discussion on the topic and  
in the role played by  $\nu$ and $\tau$ in reproducing the above relations to 
Matteucci (1994) and PM04 (see also Thomas et al.\ 2002; Ferreras \& Silk 2003).

Our model predictions for the photochemical properties related to the optical part of
the spectrum are summarized in Table 2 and compared with observation in Fig. 2 and Fig. 3.
In this section, however, we simply refer to model L (which represents models L0, La1, La10)
and model H (which represents models H0, Ha1, Ha10), since the optical properties are not
influenced by $M_{\rm acc}$.
In particular, in Table 2 we show the predicted line-strength indices (see below) in Cols. 2-5,
the [$\langle$Mg/Fe$\rangle$] 
in Col. 6 and the luminosities of the whole galaxies in the K and L-band 
in Cols. 7 and 8, respectively.
The predicted indices are evaluated by calculating the mean [Mg/H], [Fe/H] and [Mg/Fe]
in the stellar component of the galaxies, and then converted into Mg$_2$ 
and $\langle$Fe$\rangle$ by means
of a suitable \emph{calibration}. In particular,
the indices obtained by Matteucci et al.\ (1998) with 
the Tantalo et al.\ (1998) $\alpha$-enhanced SSPs are labelled 
with T, whereas W refers to the Worthey (1994) SSPs (see PM04). For a detailed discussion
on the adopted procedure and a comparison between the two sets of calibration see PM04.

\begin{table*}
\centering
\begin{minipage}{120mm}
\scriptsize
\begin{flushleft}
\caption[]{Predicted optical properties of the model galaxies}
\begin{tabular}{lllllclll}
\hline
\hline
Model & ${\rm Mg_2}$ &        & $\langle$Fe$\rangle$ &        &
[$\langle$Mg/Fe$\rangle$] & $L_B$ & $L_K$\\
      &  T     & W      & T      & W     &         &($10^{10}L_{\odot}$)&($10^{10}L_{\odot}$)\\
\hline
L    &  0.301  &  0.232  &  3.00  &   2.68    &0.506  &	1.2&	7.7\\
H  &  0.304  &  0.220&    2.93  &  2.57  &  0.561  &	11.2&	80\\
HSN1&0.431   & 0.350  &  3.94  &  3.78   &  0.462   & 18 &116 \\
HSN100&0.268   & 0.172  &  2.61  &  2.10   &  0.689   & 4.4 & 19.8\\
\hline
\end{tabular}
\end{flushleft}
The indices obtained with the Tantalo et al.\ (1998) SSP are labelled with T, whereas W refers to the Worthey (1994) SSP (see text). 
The values presented here refer to the whole galaxy.
\end{minipage}
\end{table*}

Before discussing the agreement with the observed mass-metallicity relation,
we stress that the observed ones refer to the central (i.e. $\sim 0.1
R_{\rm eff}$) galactic region,
whereas we are presenting the results for a one-zone model extending out
to $10 R_{\rm eff}$. We also recall
that the strength
of the metallic lines is different on these two scales, being weaker
at larger radii (e.g. Carollo et al.\ 1993).
Therefore, in order to compare our predictions with observations in Fig. 2, we must convert our average results for the whole galaxy
(entries of Table 2)
into a precise estimate for its central region. In particular, we derived the line-strength indices for the central zone as 
$Index_0 = Index(1R_{\rm eff}) + \Delta Index$, where $\Delta Index$ is 
the absolute value of the theoretical
gradients predicted by PM04 best model within 1 $R_{\rm eff}$. In order
to do that, we assume that $Index(1R_{\rm eff})\sim Index(10R_{\rm eff})$,
namely that the mean stellar metallicity, either measured or predicted,
at $\sim R_{\rm eff}$ is 
representative of the whole galaxy (e.g. Arimoto et al.\ 1997).
It is worth noticing that when considering line-strength indices derived by $\alpha$-enhanced tracks (solid curves in Fig. 2), 
the negative gradient in [Fe/H] is balanced by a positive radial gradient in [Mg/Fe] (see PM04), so the values for Mg$_2$ and $\langle$Fe$\rangle$
predicted by the one-zone model not only preserve the slope, but also lie very close to the average
observed ones.

Our fiducial models can reproduce reasonably well also the [Mg/Fe]-$\sigma$ relation, once the entries
in Table 2 are corrected for the radial gradient. In particular, we assume the values
given by PM04 for their case IIb, namely $\Delta$ [Mg/Fe]$=$0.326 for model L and $\Delta$ [Mg/Fe]$=$0.169
for model H, leading to the predicted central values [Mg/Fe]$=$0.180 (L) and [Mg/Fe]$=$0.392 (H), which
are in good agreement with the observations.

\begin{figure}
\includegraphics[width=\hsize]{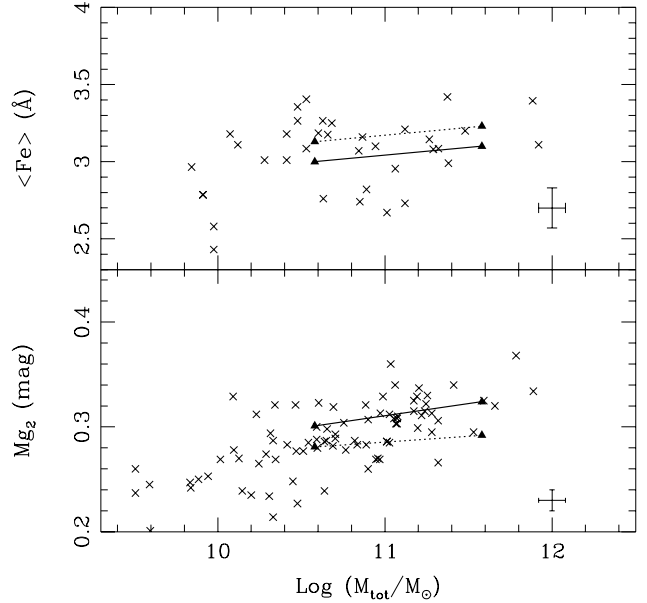}
\caption{Predicted mass-metallicity relation for the central zones of our models. 
The curves are obtained by correcting the average properties of the whole
galaxy with the theoretical metallicity gradients by PM04 (see text for details).
Results for 
$\alpha$-enhanced calibration (solid line) and not $\alpha$-enhanced calibration
(dashed line) are shown versus a collection of data from Kuntschner (2000) and Kuntschner et al.\ (2001).
}
\end{figure}

We show the predicted $V-K$ versus $M_V$ and $U-V$ versus $M_V$ relations in Fig. 3 for two different
ages for the galactic models (the fiducial case with $\epsilon_{\rm
SN}=0.1$) compared with the data of Bower et al.\ (2002b). 
In order to satisfy the condition given by the CMRs, we assume that the galaxy in model H forms 
at redshift 3 (age of 11.3 Gyr)
while in model L it forms at redshift 2.6 (age of 11 Gyr) in an
$\Omega_m=0.3$, $\Omega_{\rm \Lambda}=0.7$,
$H_o=70 \rm \,km\,  s^{-1}\, Mpc^{-1}$ cosmology. The fact that the less massive galaxies
might be also the youngest, has been recently suggested by several authors 
from the analysis of their observed line-strength indices 
(e.g. Thomas et al.\ 2002).

In order to show how the choice of the age can affect the predicted CMRs (age-metallicity degeneracy, e.g. 
O'Connel 1976), we show a case in which the galaxies in both models L and H
form at redshift 5 (i.e. they have an age of 12.3 Gyr for the assumed cosmology), with full
triangles in Fig. 3.

\begin{figure}
\includegraphics[width=\hsize]{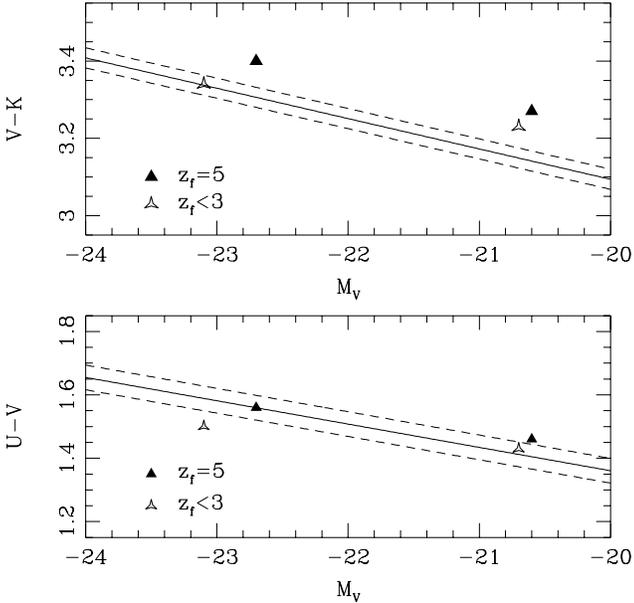}
\caption{Predicted CMRs for our models and different redshift of formation compared with
the fit to the Bower et al.\ (2002b) data (solid line). A measure of the observational scatter is given
by dashed lines.}
\end{figure}

Either models with a very low (HSN1) or maximum (HSN100) SN efficiency
fail in reproducing the above relations (Figs. 1 and 2). As expected, model HSN1 hardly develops
a wind, thus allowing SF to take place until the present time. SNe II explosion rate
inconsistent with observations (Cappellaro et al.\ 1999), too blue colours and, at the same
time, a too high stellar metallicity with respect to model H (and hence to the observed values)
are natural consequences of this particular SF history.

Model HSN100, on the contrary, undergoes a galactic wind too early, thus 
not allowing for the necessary enrichment of the stellar component up to 
$\langle Z\rangle\sim 0.02$. 
As it can be clearly seen
from the entries of Table 2, in this case the line-strength indices are too weak,
and the galaxies exhibit very low luminosities.
For these reasons, model HSN1 and HSN100 will not be analysed in further detail in the remainder of the paper.

\section{The gaseous halo - X-ray properties of the models}

\subsection{The energy evolution of the ISM}

Once we have tested our models in their photochemical properties, showing that the new energetic
formulation preserves the results achieved by PM04, 
we dedicate this section and the following one
to our new results. In particular, we first concentrate on the temporal behaviour of cooling, gas density and
temperature of the ISM, in order to understand what determines the $L_X - L_B$ and $L_X - T_X$ relations.
Then we shall analyse the chemical composition of the predicted X-ray haloes.
A summary of the predicted properties for the ISM of the model galaxies
is shown in Table 3. For each model we shall show the role played by secondary accretion
in determining the final hot gas mass (Col. 2), chemical abundances (Cols. 3 and 4), X-ray luminosity
(Col. 5) and, finally, temperature (Col. 6).

\begin{table*}
\centering
\begin{minipage}{120mm}
\scriptsize
\begin{flushleft}
\caption[]{Predicted X-ray properties of the model galaxies}
\begin{tabular}{lllllll}
\hline
\hline
Model & $M_{\rm gas}/ M_{\rm lum}$      & [Fe/H] & [O/Fe]      &$L_X$ &$\rm kT_X$\\
      &                         &        &            &($10^{41}\rm erg\, s^{-1}$)&(keV) \\
\hline
L0    & $0.14\cdot10^{-3}$        &0.85    & -0.56      & $9.3\cdot10^{-3}$ &1.2 \\
La1   &0.014                    &0.85    & -0.68       & 2             &0.5\\
La10  &0.15                     &0.92    &-1.05       & -             &0.001\\
H0  &  $0.3\cdot10^{-2}$        &1.23   & -0.87     & 16.7 &1.7 \\
Ha1   &   0.05                   & 1.3&  -1.04&    183 &1\\
Ha10  &  0.17                   & 1.60&  -1.17&    - &0.001\\
\hline
Ha1-0.3$Z_{\odot}$   &   0.05                   & 1.4&  -1.02&    175 &0.98\\
Ha1-$Z_{\odot}$   &   0.05                   & 1.3&  -0.98&    254 &0.84\\
\hline
\end{tabular}
\end{flushleft}
Models Ha1-0.3$Z_{\odot}$ and Ha1-$Z_{\odot}$ are similar to model Ha1, but
the second accretion episode has a metallicity of 0.3$Z_{\odot}$ and $Z_{\odot}$,
respectively.
\end{minipage}
\end{table*}

As it can be clearly seen from the entries of Table 3, models which accrete $\sim 0.1 M_{\rm lum}$ or more during their entire lifetime,
undergo strong cooling and exhibit too low temperatures to emit in the X-ray.
In any case, a variation in $M_{\rm acc}$
of a few percent can dramatically change the late time evolution of the galaxies.
This finding can be explained by means of Fig. 4, using model H as an example. 
First of all, it is worth noticing that the three models show no relevant differences in the heating
source, because at late times only SNe Ia are exploding and their rate is a function
of the SF history (which is the same, since the secondary accretion 
affects neither the SFR nor the galactic wind occurrence).
Model H0 undergoes galactic wind for a longer period. At every time-step the
condition $E_{\rm th}>E_{\rm Bin}$ is satisfied and this
implies $\Delta E>0$ and $\Delta M>0$, which, in turn, leads to an increase in the heating per unit mass
and a decrease in the cooling (due to the decreasing density). This mechanism
leads to slightly high temperatures ($>2$ keV) and devoids the galaxy
of most of its gas.
The opposite behaviour is showed by model Ha10, where the condition
$E_{\rm th}>E_{\rm Bin}$
cannot hold for a long time interval, owing to the higher mass accretion rate and faster
cooling with respect to model H0. 
It should be remarked that, in models Ha10 and La10, the accretion rate at $t>t_{gw}$ exceeds
the mass loss rate by a factor of 2, although $M_{acc}$ is 
much lower than $dM_w /dt$ integrated over the entire galactic lifetime.
This is due to the fact that, $\sim 100$ Myr after $t_{gw}$, only low mass stars
(i.e. mass $< 3 M_{\odot}$) are still alive in the galaxy, and their
contribution to the returned mass is quite low.  
\rm
When the galactic wind stops, all the gas either ejected by stars
or accreted from the ICM is retained and leads to an even stronger cooling. The galaxy eventually exhibits a cold
ISM. 

Models with $M_{\rm acc}\sim 0.01 M_{\rm lum}$ (i.e. La1 and Ha1), on the contrary, show a reasonable behaviour
of the gas temperature, with a good balance between heating sources and cooling term,
given the fact that the accreted gas has the virial temperature of the galaxy.
Similar conclusions were reached in previous works (e.g. Ciotti et al.\
1991; Lowenstein \& Mathews 1987; Ferreras et al.\ 2002). 
Ciotti et al.\ (1991), pointed out the importance of a
late time inflow phase in order to explain the high X-ray luminosities
of the most massive objects, although the gas accretion they considered was limited only to the innermost regions of the galaxies
and the inflowing gas had a stellar origin. From these results, in the following we focus on models La1 and Ha1.

\begin{figure}
\includegraphics[width=\hsize]{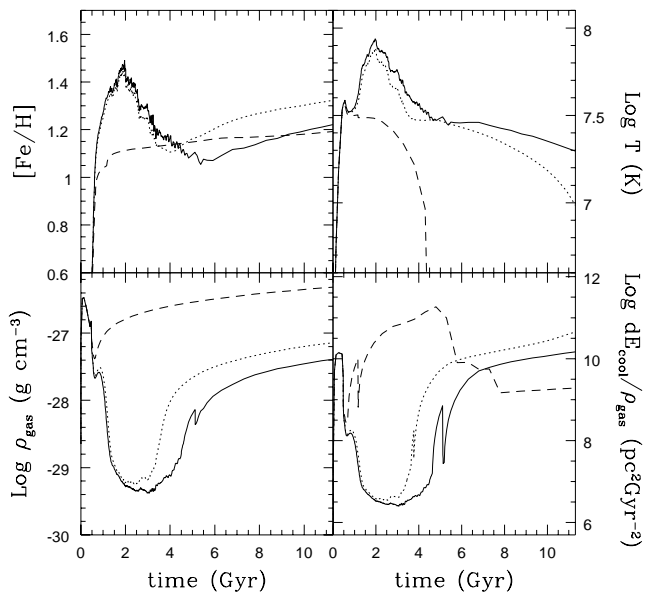}
\caption{Temporal behaviour of the ISM iron abundance, temperature, gas density and cooling
term predicted for models H0 (solid lines), Ha1 (dotted) and Ha10 (dashed).}
\end{figure}

\begin{figure}
\centering
\includegraphics[width=\hsize]{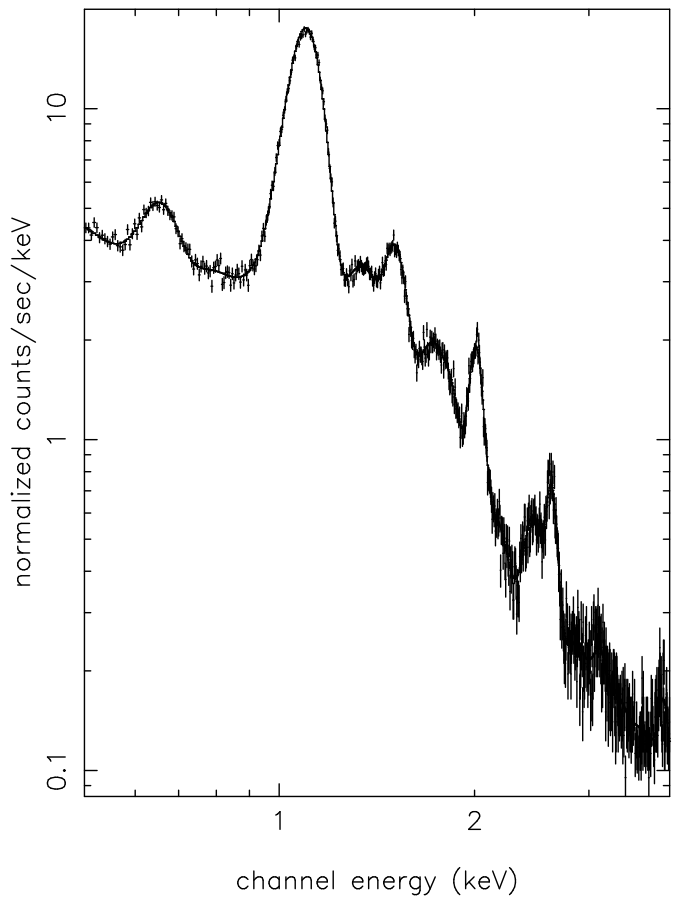}
\caption{Predicted X-ray spectrum for model Ha1. For the adopted procedure see Sec. 2.7.}
\end{figure}

In Fig. 5 we show
the spectrum predicted for model Ha1 following the procedure described
in Sec. 2.7. It is in good qualitative agreement with the observed spectra
(see e.g. Gastaldello \& Molendi 2002),
and we notice a strong feature, centered at 1 keV, due to the very high
Fe abundance predicted by our models.
Among the metal lines, in fact, the Fe L shell gives the highest
contribution, owing to the fact that the ISM of the model galaxies is 
rich in iron.
In order to have more quantitative constraints for our model (with particular emphasis
on models La1 and Ha1), we extract 
information about $L_X$, $T_X$ and [Fe/H] from the fake spectra and 
compare them with observations. 

\subsection{The $L_B-L_X$ relation}

First of all, we show how our models
can match the $L_B-L_X$ relation, which is the best determined
link between the optical and the X-ray part of the spectrum 
(either theoretically or observationally, see Sec. 1). 
In Fig. 6 we show our predictions
for model La1 and Ha1 (filled triangles), compared with the mean relation
derived from the large sample of O'Sullivan et al.\ (2001a, solid line). Our points lie well within
the observed region (a measure of a $1\, \sigma$ deviation is given by the dotted lines), although the predicted slope seems to be
slightly flatter with respect to the fitted relations.

An interesting result is that the scatter in the observed $L_B-L_X$ relation can be totally explained
either by small differences in the accretion history or variations in density (as we explain later in this section, 
see also Ciotti et al.\ 1991). In order to show this,
we plot in Fig. 6 models with $M_{\rm acc}=0.005M_{\rm lum}$ and
$M_{\rm acc}=0.007M_{\rm lum}$ (filled squares and stars, respectively)
as well as models L0 and H0 (empty squares).
The difference in the predicted $L_X$ at a given $L_B$ for the accretion models and the cases without secondary
inflow spans the same region that is between the fit to the observed data (solid line in Fig. 8) and the 1$\sigma$ boundary (dotted line)
in the case of the high mass model. For the low mass one, instead, the range in the predicted values covers
a much larger region. 
This fact is a consequence of the adopted formalism for the energetics. In fact, with the exception
of the very beginning of the galactic wind, when $E_{\rm th}\gg E_{\rm Bin}$, the condition on the
mass flow presented in eq. (14) coupled with a mild accretion of slightly colder external gas,
implies that $E_{\rm th}\geq E_{\rm Bin}$ at each time step.
In this cases, the mass accretion rate is 0.2 - 0.5 times the value
of the mass loss rate. Therefore it gives a non-negligible contribution
to the building-up of the X-Ray emitting ISM. 
\rm
Even small changes in the accretion 
rate can cause differences in the predicted properties. 
The higher density of the low mass model with respect to model H amplifies this effect.
In this case, galaxies with relatively large optical luminosities but
very low $L_X$ (e.g. the recent detection
by O'Sullivan \& Ponman 2004) can be explained as objects still in the process of assembling the
X-ray emitting ISM. Observations (e.g. O'Sullivan et al.\ 2001b; 
Samsom et al.\ 2000)
showing a correlation linking the age of the galaxies to their $L_X$, in the sense
that younger systems exhibit lower luminosities, give further support to this view.

On the other hand, we cannot present more quantitative results, because of the number of parameters involved.
The assumed radius, in fact, can play an important role as well, because the change of the dimension
of the system leads to a change in the mean density and, thus, strongly affects the cooling (see Sec. 2.2.3). 
In Fig 6, the prediction for model Ha1 with $R=8R_{\rm eff}$
is plotted (empty hexagon). This galaxy exhibits a very high $L_X$, when compared with other models at nearly the same $L_B$,
owing to its stronger cooling which allows more gas retention. When
smaller radii (e.g. $R=3R_{\rm eff}$) are assumed, however, 
the cooling becomes so strong that the galactic wind is delayed and lasts for a shorter time than in
model Ha1. Therefore, the models undergo a late time behaviour similar to
model Ha10, namely an ISM gas mass of $\sim 0.1-0.2M_{\rm lum}$,
very low temperatures ($\sim 10^4$ keV), and [Fe/H]$>1.2$, clearly at variance with observations.
Since these models were run keeping all the other parameters fixed only for comparative purposes, they
are not meant to match the optical properties and will not be discussed further.

The natural conclusion is that the late accretion can play a role in the present-day $L_B-L_X$ relation, its scatter being
possibly related to the differences in the gas reservoir from which a galaxy can form 
and possible different histories and efficiencies governing 
the secondary accretion.

We anticipate from Sect. 5.4 that our model exhibits [Fe/H] higher than solar, at odds with observations.
Therefore, in order to check whether a possible strong contribution from iron line emission can alter our results,
we modelled two additional spectra by forcing [Fe/H]=0 in the ISM of both galaxies La1 and Ha1. The luminosities
obtained from these spectra (shown by empty triangles in Fig. 6) are lower than those emitted by La1 and Ha1, and the differences
are of the same order of those produced by a change in the accreted mass.

\rm
In order to test the assumption of primordial composition for the second accretion flow,
we run models Ha1 in which the accreted gas has $Z=Z_{\odot}$ and $Z=0.3\, Z_{\odot}$, but always
solar abundance ratios. From the entries in Table 3 it is clear that
our results are robust against changes in the composition
of the accreted gas.
We did not comment cases with a $Z=0.3-1\cdot Z_{\odot}$ metallicity and a higher accretion rate, because the model predictions
still show too cold galaxies to emitt in the X-Rays.
\rm

\begin{figure}
\includegraphics[width=\hsize]{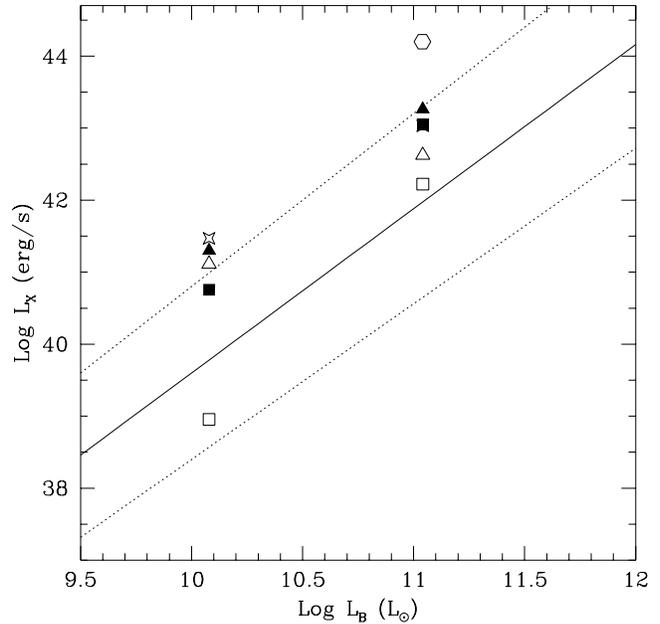}
\caption{
Prediction for the $L_B-L_X$ relation for our model galaxies compared 
to the best fit of O'Sullivan et al.'s (2001a, solid line) data, 
excluding bright cluster and group ellipticals as well as galaxies hosting an AGN. 
The dotted lines give a measure of the scatter. Filled triangles: La1, Ha1. Open triangles: La1, Ha1 in which we fix [Fe/H]=0.
Stars: Models L and H with $M_{\rm acc}=0.007M_{\rm lum}$. Filled
 squares: Models L and H with $M_{\rm acc}=0.005M_{\rm lum}$. Open
 squares: H0 and L0. Hexagon: Ha1 with R=8$R_{\rm eff}$.
}
\end{figure}

\subsection{The $L_X-T_X$ relation}

In Fig. 7 we compare our predictions in the $L_X-T_X$ plane with the data by Matsushita et al.\ (2000).
In this particular case we show $L_X$ in the passband 0.5-10 keV, in order to be consistent
with the observed energy range. The agreement is good for the La1 and Ha1 models, whereas
the ones without accretion exhibit a slightly higher temperature. In any case,
a factor of two in the temperature can be partly explained with
variation in gas density and accretion history, but also might be related to the adopted one-zone formalism. 
Ferreras et al.\ (2002) presented a similar model for feedback and found a good agreement for the $L_B-L_X$ relation,
even though their low mass model exhibits a temperature of a few keV.
They did not model any interaction with the environment, and thus any secondary accretion; moreover the
fraction of gas allowed to escape was chosen a priori. This last point may represent another
way to regulate the temperature in the ISM, without requiring secondary accretion.
In fact, if we do not link
the mass flow to the energetic budget and we have a situation in which 
$\Delta M/ M_{\rm gas}<\Delta E/ E_{\rm th}$, the result is that the galactic wind removes
the gas inhabiting the high energy tail of the temperature distribution function,
thus allowing the galaxy to retain more gas and eject more energy with respect 
to the case with no accretion. Therefore, the heating term cannot increase without limits
even in presence of a continuous wind. 

We did not further explore this route, since it introduces an additional parameter
to our formalism, but we stress that it might represent a natural solution
in a more detailed modelling (e.g. two-phase ISM).

A better agreement is achieved when we consider the models La1 and Ha1 in which
a [Fe/H]=0 ISM abundance ratio is forced (empty triangles in Fig. 7).

\begin{figure}
\includegraphics[width=\hsize]{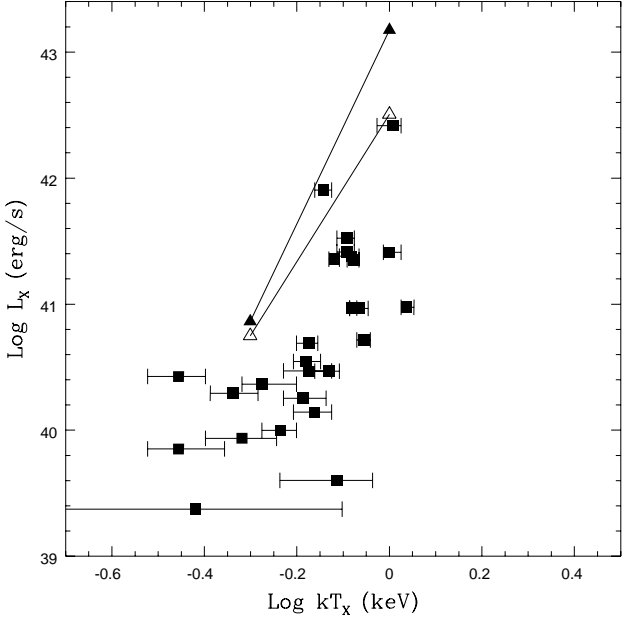}
\caption{Prediction for $L_X-T_X$ relation for our model galaxies La1
and Ha1 (filled triangles) compared to the data of
Matsushita et al.\ (2000, squares with error bars).
For comparison we show the models La1 and Ha1 in which we fix [Fe/H]=0
(open triangles).}
\end{figure}

\subsection{The chemistry of the ISM}

Concerning the study of the chemical composition of the ISM, we
first analyse the iron abundance, which is the best determined
among all the metals.
Predictions for the $L_X-{\rm [Fe/H]}_X$ relation for our model galaxies La1 and Ha1 compared to
Matsushita et al.\ (2000, squares) data are shown in Fig. 8 with filled triangles.
As expected from the analysis of the fake spectrum, our models
predict a very high iron abundance, even if we compare our results
with recent iron abundance determination, going from the typical 
$\sim 1-2\, {\rm Fe_{\odot}}$ in the central region
of giant elliptical galaxies (e.g. Xu et al.\ 2002; Buote 2002)
up to the $\sim 3-4\, {\rm Fe_{\odot}}$ in the case of NGC507 (Kim \& Fabbiano 2004).
In the latter case, however, our model predictions are only within a factor of 2 from
the observations. 

This high abundance is a clear consequence of our model, 
since all the iron produced,
after the wind has stopped, is retained by the galaxies. 
In other words, the iron discrepancy seems to be still persisting.

\begin{figure}
 \includegraphics[width=\hsize]{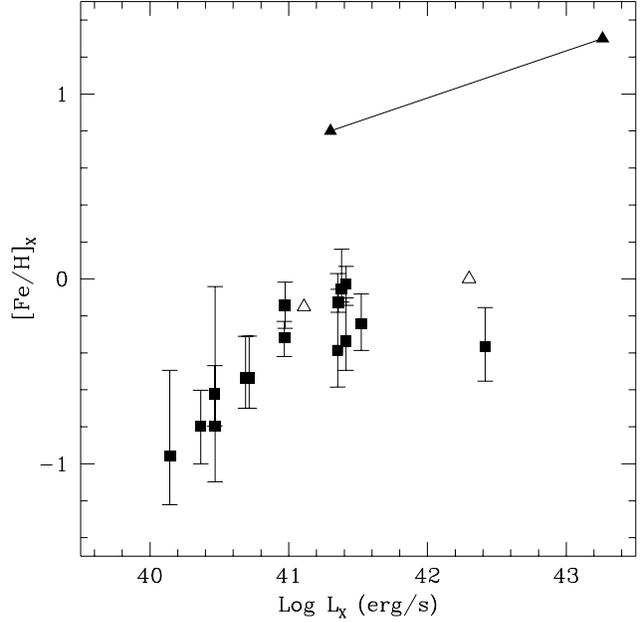}
\caption{Prediction for $L_X-{\rm [Fe/H]}_X$ relation for our model
galaxies La1 and Ha1 compared to the data of Matsushita et al.\ (2000).
 The symbols are the same as in Fig. 7. Empty triangles
refer to models Ha1 and La1 with the same $L_X$, but with [Fe/H]$\sim 0$.}
\end{figure}

Among the possible physical mechanisms often invoked to solve this discrepancy (see
Arimoto et al.\ 1997 for a comprehensive analysis), we recall: (i) differential winds
powered by SNe Ia and (ii) iron hidden in a colder phase. According to the former explanation,
although our models stop the \emph{global} wind phase, SNe Ia ejecta might be still able to escape
the galactic potential well, thus lowering the iron abundance in the ISM (e.g. Recchi et al.\ 2001).
In order to preserve the observed $L_X - L_B$ and $L_X - T_X$ relations,
this could be obtained by e.g. requiring that $\Delta M \sim M_{\rm Fe}^{ISM}$ holds at each time-step, namely
only the iron can escape, whereas the bulk of the gas is retained. This, in turn, implies that
$\Delta M/ M_{\rm gas}<\Delta E/ E_{\rm th}$.

The latter possibility, on the other hand, explains the low observed [Fe/H] by means of condensation
in a colder phase (e.g. Fujita et al.\ 1996, 1997) or in dust grains. Recent observations,
in fact, claim that the dust mass in ellipticals could be $\sim 10^{6-7}M_{\odot}$,
a factor of ten higher than previous estimates (Temi et al.\ 2004). If confirmed,
this result will be at variance with the classical argument that
the dust lifetime is short owing to the interaction with the hot ISM, and therefore it might represent
a viable solution to hide huge amounts of iron for a long period (Arimoto et al.\ 1997).

However, recent observations (e.g. Bohringer et al.\ 1993; 
Finoguenov \& Jones 2001; McNamara et al.\ 2000; Allen et al.\ 2001;
Heintz et al.\ 2002) 
show that both the X-ray haloes
and the region of the ICM very close to the central galaxy, are not uniform
and that the presence of X-ray cavities, bubbles with relativistic gas that pushes away
the `thermal' gas, could be related to the AGN activity 
(Churazov et al.\ 2001). 
Central emission and the creation of radio lobes can drive a mixing between the ICM and the ISM in such a way that part of the Fe
is removed from the central galactic region. If we allow for the excess mass of iron to be
replaced by primordial gas, our models do not change their ISM gas mass, thus roughly preserving their $L_X$, 
whereas the iron abundance is reduced to the observed values (empty triangles in Fig. 8). For example, 
in the simplistic case in which the metallicity is composed only of iron,
if we assume that an AGN engine causes mixing to take place from
the time at which the galactic wind stopped until the present day, 
and we want to remove the iron mass in excess ($\sim M_{\rm Fe}^{ISM}$)
in order to end up with [Fe/H]$=0$ in the ISM, 
the rate of mass exchange is only 
$\sim M_{\rm Fe}^{ISM}/ 5 Gyr \sim 0.5\rm M_{\sun}yr^{-1}$. 
In the case of mixing involving non-primordial gas 
(e.g. [Fe/H]$\sim-0.5$ in the ICM), instead, it should be 
either more extended in time or require a slightly higher rate of mass exchange.
Moreover, an additional source of heating (such as an AGN) might balance the strong cooling in model Ha10, 
thus allowing the galaxy to maintain a longer wind phase. This, in turn, implies
that more iron could be injected into the ICM and that the iron still present in the halo
is diluted by a higher amount of gas with respect to our fiducial case (model Ha1). At the same
time the galaxy could exhibit the right temperature and gas mass in order to match the $L_X - L_B$
relation. Moreover, the presence of an AGN could supply the amount of energy 
needed to mantain the galactic wind, in a scenario in which we onsider $\epsilon_{\rm SN Ia}=0.1$
fixed for the whole galactic lifetime.
In concluding, we note that evidences on the non-stellar origin of either a part of or all the ISM
are given by the observation of subsolar iron abundance 
in galaxies which lie well
below the average relation in the $L_X - L_B$ plane (e.g. O'Sullivan \& Ponman 2003), although
the poor statistics might bias the observations (Matsushita et al.\ 2000).
Further aspects of this issue will be investigated in the future.

Another possible solution can be a variation either with time or metallicity of the fraction 
of binary systems (e.g. De Donder \& Vanbeveren 2002). This can be implemented in our code 
by assuming that the parameter $A$ in the definition of SNe Ia 
rate (Matteucci \& Greggio 1986) is not constant in time.
We tested a model in which $A$ scales linearly from 0 to the assumed value of 0.09 with the gas metallicity, until the 
solar value (Z=0.02) is reached, and we did not find any significant reduction
in the SNe Ia rate and the late time iron production. This is due to the very intense
SF history experienced by ellipticals, which leads to a very fast increase
of Z to the solar value in the first $\sim 100$Myr of evolution.

Concerning other chemical species, they represent a weaker constraint
to our modelling than Fe, owing to the large observational uncertainties
still affecting their determinations.
A sample of recent observations is shown in Table 4.
The galaxy identity is given in Col. 1. 
Chemical abundances are presented in Cols. 2-9. In particular,
subscript \emph{c} means values measured in the core of the galaxy, whose radius
is given in Col. 10, whereas subscript \emph{o} is related to properties
of the outskirts. Finally, Cols. 11 and 12 show the adopted solar composition
and the references.

\begin{table*}
\centering
\begin{minipage}{200mm}
\scriptsize
\begin{flushleft}
\caption[]{Chemical abundances in ellipticals from their X-ray spectra}
\begin{tabular}{llllllllllll}
\hline
\hline
Galaxy	&             [Fe/H]   &[O/Fe]&	[Mg/Fe]&[S/Fe]&	[Si/Fe]& [Ne/Fe]& [C/Fe]&  [N/Fe]&	core&	ab.&  ref.\\
&&&&&&&&& (kpc) &  &\\
\hline
NGC5044	& 		$\sim 0_c/-0.36_o$ &	$-0.48$&	$-0.06$&	$-0.27$&	$-0.08$&	-&  -	&  -  &    48&	GS &$^a$ \\
M87	&		$-0.11$&	   	$-0.44$&	$-0.15$&	-&	-&	$-0.26$&	  0&	  $-0.21$  &  -     &        AG & $^b$\\
M87&		-	&		    $-0.52_c/-0.28_o$ &	$-0.28$&   $-0.08$&	$0_c/0.1_o$&$-0.52_c/-0.21_o$&  - &   -&      2.5&	GS& $^c$\\
M87&                    $-0.009_c/-0.39_o$  & $-0.34_c/-0.20_o$  &   -  &$0_c/-0.09_o$   &$0.05_c/-0.04_o$ &-&  - &  -  &           - & F & $^f$\\
NGC1399	&		$\sim 0.3_c/\sim -0.3_o$  &     -&	-  &  -&	-&	-&	   - &  -&       20 &    AGm&$^d$\\
NGC4636	&		$-0.06$&			$-0.24$&	$-0.13$&  -&	-&	$-0.09$	&   -&	  0.07	 &    - &	AG & $^e$\\	  
\hline
\end{tabular}
\end{flushleft}
Solar abundance used: AG: Anders \& Grevesse (1989, photospheric), 
AGm: Anders \& Grevesse (1989, meteoritic), GS: Grevesse \& Sauvall (1998), \\
F: Feldman (1992). \\
References: $^a$: Buote et al.\ (2003), $^b$: Sakelliou et al.\ (2002),
$^c$: Gastaldello \& Molendi (2002), $^d$:  Buote (2002), $^e$: Xu et al.\ (2002), $^f$: Matsushita et al.\ (2003).\\
Core: radius of the region quoted as `galactic core' in the above-mentioned paper. 
\end{minipage}
\end{table*}

As it can be noticed from Table 4, there is a lot of confusion
created by different adopted solar compositions. In the following we present our
results adopting the solar meteoritic values of Anders $\&$ Grevesse (1989).
Even though the presence of radial gradients in the [$\alpha$/Fe] abundance ratios
is still debated (e.g. Gastaledello \& Molendi 2002;
 Matsushita et al.\ 1997), the observations show subsolar ratios, 
especially in the cases of [O/Fe] and [Mg/Fe].
In our predictions, [Mg/Fe] traces very well [O/Fe], being both $\sim-0.6$ dex, whereas
Si and Ca show a milder depletion ($\sim-0.3$ dex) with respect to the solar level.
In any case, these predictions are $\sim-0.2$ dex lower than the measured values in the central (i.e. $\sim 50$kpc)
regions of X-ray haloes, although still within their quoted $\sim 1-2\sigma$.

On the other hand, the predictions from models undergoing a continuous wind (L0 see Table 3)
are in better agreement with the observations presented in Table 4. It is worth noticing
that these predictions reflect the high [$\alpha$/Fe] ratios typical of the ejecta
of low mass stars formed at the beginning of the galactic evolution and dying after 10 Gyr.
This implies that, in order to reconcile our predictions with
observations, a mechanism capable of ejecting iron in a more efficient 
way with respect to other elements might be better than
a simple dilution by primordial gas, which preserves the abundance ratios.

We stress that the disagreement can be partly alleviated (at least at larger radii) by means of a multi-zone
formulation, which takes into account the fact that at radii $>1R_{\rm eff}$
the SF lasts for a much shorter timescale with respect to the galactic
core, thus reducing the iron production and exhibiting much higher [$\alpha$/Fe] ratios
in the ejecta of low mass stars dying in the late stages of the galactic 
evolution.
Therefore, more detailed modelling is required to assess this issue and to check
whether the observed gradients can be reproduced.

The observed ratios and the different behaviours shown by $\alpha$-elements
(and similar abundance patterns observed in the ICM, see next section) were used in 
recent works (Finoguenov et al.\ 2002, Gastaldello \& Molendi 2002) in order to infer information on a possible
change in the SNe II/SNe Ia ratio as well as on different explosion mechanisms
as a function of galactic radius. Our opinion is that this kind
of diagnostic is still premature, given the uncertainties in the models
used to fit the spectra (Mathews \& Brighenti 2003; Gastaldello \& Molendi 2002).
Moreover, we stress that it is incorrect
to draw firm conclusions from observed abundance patterns 
just by relating abundance ratios to stellar yields and thus implicitly assuming
the instantaneous recycling approximation.

\section{The ICM}

\begin{figure}
\includegraphics[width=\hsize]{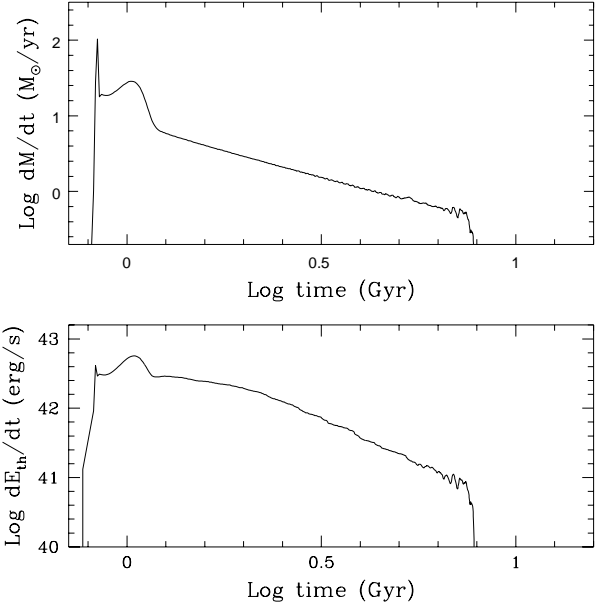}
\caption{Mass and energy flow rate versus time during the galactic wind
for a La1 model galaxy.}
\end{figure}

Before concluding, we dedicate a section to the chemical enrichment
of the ICM, since the improved chemical formalism presented in this paper allows us to follow
the development of the galactic wind in a more detailed way with respect
to previous works (e.g. Matteucci \& Gibson 1995; Martinelli et al.\
2000; P02). In Fig 9 we show,
taking model La1 as an example, the temporal behaviour of both mass and energy flow rate.
After a very short period in which they exhibit quite high values, most of the wind
phase is governed by their slow decrease as a consequence of the secondary gas accretion and the
secular decline of SNe Ia. The mass flow rate has the same order of magnitude (in general within a factor of 2) 
of the value taken by the SFR just before the galactic wind, and this result is in agreement with 
the indications coming from the observations of starburst galaxies (e.g. Heckman 2002).
On the other hand, the luminosity of the wind, after the very first intense blowout, is $\sim 10^{41}\rm\, erg \,s^{-1}$,
giving a wind speed $v\sim (2\, dE/dM)^{1/2}\simeq 200\ {\rm km\
s^{-1}}$, again in agreement with the 
the typical estimates for starbursts (Heckman 2002).
It has been claimed (Moretti et al.\ 2003) that, in the case of
a Salpeter (1955) IMF, the SNe Ia-driven winds can last for a long period, in order to inject into the ICM a suitable
amount of iron, only in the case of very high SN thermalization efficiency
(i.e. $\epsilon_{\rm SN Ia}=1$, P02).
This is not evident, since in this paper we show that 
the wind phase is about 1/3 of the galactic lifetime, being fully supported by SNe Ia activity, even 
if $\epsilon_{\rm SN Ia}\ll \epsilon_{\rm SN Ia,P02}=1$ for most of the galactic evolution.

Assuming that the metals observed in the ICM come from elliptical (and perhaps S0) galaxies (Arnaud et al.\ 1992)
and following the same procedure as described by P02,
we found no significant changes in the predicted amounts of gas, iron, 
oxygen and energy ejected
into the ICM by the model galaxies, with respect to the old model.
In particular, the total amount of oxygen ejected into the ICM, 
$0.43\times 10^9M_{\odot}$ ($0.44\times 10^{10}M_{\odot}$)
by model La1 (Ha1), is very similar to the entries of P02 Table 1, whereas the new models
lose $0.28\times10^9M_{\odot}$ ($0.23\times10^{10}M_{\odot}$) of iron, 
which is a factor
of 2 higher with respect to P02 model I (but in agreement with
their estimate for model II), owing to the more extended SF history which has increased
the amount of iron released promptly at the beginning of the wind. 
Therefore, even though the new models predict ellipticals slightly younger than those in previous works,
the amount of iron in the ICM can still be explained by means of the galactic winds,
and we can recover the typical $\sim 0.3\, Fe_{\odot}$ (e.g. recent review by Renzini 2003; White 2000)
by assuming that most of the gas in the ICM has a primordial origin (Matteucci \& Vettolani 1988,
David et al.\ 1991, Renzini et al.\ 1993, P02). Moreover, the bulk of iron is ejected before at z $>$ 1, in agreement
with recent observations on high redshift galaxy clusters (Tozzi et al.\
2003). The amount of energy 
ejected by our model galaxies into the ICM, is simply 
the integral of the curve in the lower panel of Fig. 9. 
The results are very close to
those derived by P02 for their Model I, therefore still not sufficient
to provide the required $\sim 1$keV in order to break the cluster self-similarity (e.g. Tozzi \& Norman 2001;
Loewenstein 2001; Bialek et al.\ 2001; Borgani et al.\ 2001).

Concerning the oxygen and other $\alpha$-elements, new data appeared in the last few years (see the recent review by
Loewenstein 2003) and a more consistent picture seems to be emerging from those measurements.
In particular, more accurate [O/Fe] ratios are observed for a larger sample of galaxy cluster (Peterson et al.\ 2003)
and, added to previous detections (e.g. Tamura et al.\ 2001), seem to point toward generally subsolar values.
The typical values lie around [O/Fe]$\sim - 0.2$ (e.g. Peterson et al.\ 2003), but the dispersion
is quite high, so that our predictions ([O/Fe]$\sim - 0.6$) are within $\sim 2\sigma$ (given
the typical uncertainty quoted by Peterson et al.\ 2003). Therefore, we stress that the 
new observations allowed by \emph{XMM} and \emph{Chandra} satellites seem to converge
toward a better agreement with our predictions.
On the other hand, other $\alpha$-elements exhibit different degrees of enhancement
with respect to iron (Baumgartner et al.\ 2004), although the uncertainties
associated with their observations are still very high. In particular, our
predictions for Ca ([Ca/Fe]$\sim - 0.4$) agree with the values measured by Baumgartner et al.\ (2004), 
whereas we find [Mg/Fe]$\sim -0.7$, which is at variance with the typical [Mg/Fe]$\sim 0$ (e.g. Ishimaru \& Arimoto 1997; 
Peterson et al.\ 2003, but note that their $1\sigma$ error is $\sim$0.4 dex). 
In any case a larger sample of clusters is required before drawing strong conclusions also on $\alpha$-elements. 
Furthermore, a new kind of theoretical modelling should be developed, since in many
cases the abundance ratios as well as the temperatures of the X-ray haloes of bright central galaxies  
seem to vary with radius until they reach the typical values of the ICM.
For example Peterson et al.\ (2003) explicitly noticed that their results resemble those of Xu et al.\ (2002)
for NGC4636. At the same time, the galaxies located at the center of galaxy clusters, show properties that
seem to correlate with the excess of iron measured in the center of cold core clusters and the ICM temperature (De Grandi et al.\ 2004).

\rm
Finally, we stress that if the mechanisms invoked to solve the Fe discrepancy in the ISM
were at work, our conclusions on the ICM would not have changed. For istance, the Fe \emph{excess}
in the ISM which should be removed from the halo in order to achieve
a solar abundance in the X-ray spectrum is $0.29\times 10^9M_{\odot}$ for model Ha1.
Therefore the mass of Fe expelled in this way is only the 10\% of the total mass of
Fe already ejected into the ICM during the galactic wind, and also the effects
on the predicted ICM abundance ratios will be negligible. Among the suggested mechanisms
the presence of dust could be relevant and since different scenarios could be at work together,
we believe that the real effect of Fe removal from the ISM to the ICM, will be even lower. 
\rm

\section{Conclusions}

The model presented in this paper is a first
step in the self-consistent study of both optical and X-ray properties of elliptical galaxies
by means of a chemical evolution code.
In order to do that we updated previous chemical evolution codes (PM04) by implementing
a new energetic treatment which takes into account for the first time cooling and heating processes occurring in the ISM.
Adopting the procedure described by Kawata \& Gibson (2003b) we are able to predict 
the X-ray spectrum of our model galaxies taking into account the presence of line emission
and the actual energy range used in the observations.
Here we present a summary of our main conclusions:

\begin{itemize}
\item Our new models confirm the previous finding of PM04, namely that 
SF and infall timescales decreasing with galactic mass are needed to explain the optical properties
of elliptical galaxies. 

\item At the same time we reproduce the $L_X - L_B$ relation in the ISM of bright ellipticals.
Our model can also predict the right ISM temperature in cases in which a mild
accretion from the surroundings occurs, and within a factor of 2 from the 
observed temperature at a given $L_X$ for models without accretion.
The synthesised X-ray spectrum is in a good qualitative agreement with the typical observed ones.

\item In order to model in a more realistic way the formation history of ellipticals,
we implemented the possibility of a secondary accretion episode
spread over the entire galactic lifetime. This mild inflow seems to play a non-negligible role
in governing the late-time behaviour of the ISM, in particular in driving $L_X - L_B$ and $L_X - T_X$ relations,
whereas its effects on the colours and the line-strength indices are negligible.

\item Variations in the ISM density (adopted radius) and the amount of mass accreted during the secondary infall
might account for the large spread in the observed $L_X - L_B$ relation.

\item Our model predicts [$\alpha$/Fe] abundance ratios in the ISM which are in agreement
with the data coming from very recent observations with $XMM$ and $Chandra$ satellites.

\item The \emph{iron discrepancy} between model predictions and observations still persists,
probably indicating the presence of other mechanisms acting at galactic scales. We suggest that the
mixing of gas driven by AGNs can preserve the gas mass (and thus the X-ray luminosity)
while diluting the iron abundance. 

\item The new energy formalism implemented in the chemical evolution code allows us to follow in a 
more detailed way the evolution of mass and energy flow
into the ICM with respect to previous work (e.g. Matteucci \& Gibson 1995, Martinelli et al.\ 1998, P02).
Nevertheless the predicted amount of Fe, [$\alpha$/Fe] ratios as well as the energy injected into
the ICM are very similar to P02 results. Therefore, we confirm that SNe Ia
are fundamental in providing energy and iron to the ICM and we notice that very recent observations
have reduced the gap between our predicted subsolar [$\alpha$/Fe] and the measured ones.
Even though our fiducial case has $\epsilon_{\rm SN Ia}<$1, the SNe Ia activity can power a
galactic wind lasting for a considerable fraction of the galactic lifetime and reproduce 
the observed iron abundance in clusters.

\end{itemize}

It is important to stress that we achieved these results in the assumption
of a one-zone formalism with a single gas phase in
which the model galaxy is treated like an object with uniform density.
Future developments of this kind of models should include the 
gas density profile, the possible presence of an AGN and a multi-phase 
approach. In this way we will obtain a more reliable estimate of 
the ISM temperature
and X-ray luminosity, which are strongly related to the gas profile, and
we might be able to model the gradient in the abundance ratios 
observed in both the ISM and the region of the ICM close to the brightest cluster galaxies.

\section*{Acknowledgments} 

A.P. warmly thanks the Centre for Astrophysics and Supercomputing 
of the Swinburne University of Technology for the kind hospitality in the period during which this work has been
developed. The work was supported by Australian Research Council under 
the Linkage International Award and Discovery Project schemes
and by MIUR under COFIN03 prot. 2003028039.
We thank the referee M. Loewenstein for his suggestions, which
improved the quality of the paper.
Useful discussions with S. Borgani, L. Ciotti, D. Forbes, R. Proctor, P. Tozzi are acknowledged.
We thank T. Connors for his helpful advice during the completion of this
manuscript. Finally A.P. thanks S. Recchi for many fruitful and enlightening comments.

\label{lastpage}

\end{document}